\documentstyle[12pt]{article}
\def\bcc{\begin{center}} \def\ecc{\end{center}}
\def\beq{\begin{equation}} \def\eeq{\end{equation}}  
    \def\vf{\varphi}
   
\def\la{\langle} \def\ra{\rangle} \def\r#1{$^{[#1]}$}
\begin{document}
\vskip-1.5cm
\hskip12cm{\large HZPP-9802}

%%%%%%%%%%%%%%%%%%%%%%%%% ? %%
\hskip12cm{\large Jan 25, 1998}
%%%%%%%%%%%%%%%%% ????? %%%%%%
\vskip1cm
 \date{}
 \centerline{\huge An Algorithm for Selecting QGP Candidate Events}
  \centerline{\huge from Relativistic Heavy Ion Collision Data 
  Sample$^{*}$}
\bigskip
\bigskip

 \centerline{Liu Lianshou \ \ \ \ \ Chen Qinghua\ \ \ \ \ Hu Yuan}
\bigskip

 \centerline{\small (Institute of particle physics, Hua-Zhong Normal 
University, Wuhan 430079, China)}

\begin{center} \begin{minipage}{125mm} \vskip 0.5in
\begin{center}{\bf Abstract }\end{center}
{The formation of quark-gluon plasma (QGP) in relativistic heavy ion collision,
is expected to be accompanied by a background of ordinary collision 
events without phase transition. In this short note an algorithm is proposed
to select the QGP candidate events from the whole event sample. This 
algorithm is based on a simple geometrical consideration together with some 
ordinary QGP signal, e.g. the increasing of $K/\pi$ ratio.  The efficiency 
of this algorithm in raising the 'signal/noise ratio' of QGP events in the
selected sub-sample is shown explicitly by using Monte-Carlo simulation.}

Signal of QGP formation  \ \hskip3.2cm  Factorial moments  

\end{minipage}
\end{center}
\vskip1.5cm

\vfill
 \noindent{* This work is supported in part by the NNSF of China}
\vskip1truecm

\newpage

Recently the lead-lead collision at incident energy as high as
160 A GeV has been performed at CERN-SPS. Within a few years gold-gold 
and lead-lead collsions at even higher energies  will be realized in
Brookhaven (RHIC) and CERN (LHC). At such ultra-relativistic energies
the condition for the formation of a new state of matter --- quark-gluon
plasma (QGP) will most probably be achieved. In this case, our main
concern will be turned from the question: ``Whether QGP is formed'' to the
question: ``How to identify the QGP events from the background''.  

Owing to the highly complexity of ultra-relativistic heavy ion collision, 
it could be  %% is usually 
expected that even when 
the energy threshold for QGP phase transition is reached %% formed 
in an experiment, 
the thermalization and equllibrium phase transition may not be realized 
in every event.
Complicated nuclear collision events without phase 
transition may still be present, forming a 
%% large 
background for QGP formation. 
Any physical characteristic quantity, that may serve as QGP signal, after 
averaging over the whole event sample, will 
at least partly  %%  strongly 
be smeared out. Even 
if we measure this quantity event by event and study their distribution,  
the 
%% strong 
peak from background events may 
disturb the observation of %% dominate, making 
the peak from QGP events. %% hard to observe. 
Therfore, it is worthwhile to find some effective 
way for selecting the QGP events from the whole ultra-relativistic heavy 
ion collision data sample. The aim of this letter is to propose a possible
algorithm for this purpose.

For concreteness, let us take the increasing of $K/\pi$ ratio as an example
of QGP signal. As a crude estimate\r{1} the QGP events might exhibite an
almost twofold enhancement of the average $K/\pi$ ratio $\la R_{K/\pi}\ra$ 
at mid-rapidity from the value of 0.14 for the %% ordinary 
non-QGP events. 
The relative width $\sigma(R_{K/\pi})/\la R_{K/\pi}\ra$ of $K/\pi$ ratio
distribution is estimated to be about 0.3 for CERN-SPS energy and higher. 
Let us assume the $R_{K/\pi}$ distributions for both 'non-QGP'
and 'QGP' events be Gaussian with parameters:
 $$ \la R_{K/\pi}\ra^{\rm non-QGP}=0.14 , \ \ 
 \la R_{K/\pi}\ra^{\rm QGP}=0.25, \ \  
            \sigma(R_{K/\pi})/\la R_{K/\pi}\ra = 0.3 \ . \eqno(1)$$
The corresponding distributions of 'non-QGP', 'QGP' and mixed events
for three typical cases, in which the 
percentage of QGP events $\rho_{\rm QGP}$ equal to 20\%, 33\% and 50\% 
respectively, are shown in Fig.1.

It can be seen from the figure that the shape of $K/\pi$-ratio distribution
for mixed events
is mainly determined by the non-QGP events. Even when $\rho_{\rm QGP}=50\%$ 
the QGP events only appear as a shoulder in the distribution. This shoulder can 
be ultilized in constructing a 'QGP event sub-sample'. Introducing a cut at 
$R_{\rm cut}=0.22$ and picking out the events with $ R_{K/\pi} >  R_{\rm cut}$,
a 'QGP event sub-sample' is obtained. It raises the percentage of QGP events 
from 50\% to 95.5\%, with the cost of losing 33.6\% of QGP events. 

However, when the percentage of QGP events $\rho_{\rm QGP}$ is as low as
33\% or 20\%, no evident shoulder can be seen and the method of using $K/\pi$
ratio to select QGP events failed.

In order to increase the efficiency of QGP event selection, we make use of
a simple geometrical consideration. 
As is well known, heavy ion collision at conventional energies can be regarded
as the superposition of a large number of elementary collision processes. 
A necessary condition for the formation of QGP is the thermalization of
the produced particles in individual elementary collision processes, formating
a unified system. Therefore, if we could get a criterion to judge whether the
final state particles are coming from a unique system or from a large number
of sub-systems, we would be able to make use of this cretirion to increase the
purity of 'QGP event sub-sample'.

Such a criterion can be provided by the higher-dimensional factorial moment 
(FM) analysis\r{2,3}.  The $q$th order factorial moment $F_q$ is defined 
as\r{4}
   $$ F_q(M)={\frac {1}{M}}\sum\limits_{m=1}^{M}{{\langle n_m(n_m-1)\cdots
      (n_m-q+1)\rangle }\over {{\langle n_m \rangle}^q}}, \eqno(2)$$
where a region $\Delta$ of phase space is divided into $M$ cells, $n_m$  is 
the multiplicity in the $m$th cell, and $\langle\cdots\rangle$ denotes 
vertically averaging over many events.
It has been shown\r{2} that in the 2-D($\eta$, $\vf$ or $y$, $\vf$) and 3-D
($\eta$, $p_t$, $\vf$ or $y$, $p_t$, $\vf$) space, due to the superposition 
of particles coming from a large number of sub-systems, the log-log plot of 
FM versus $M$ for nucleus-nucleus collisions will be strongly bending 
upwards.  On the other hand, if there is only one unique system 
the upward-bending of ln-FM versus ln-$M$ will disappear.

Basing on the above observation, we propose the following algorithm for
selecting QGP events from the whole data sample.

(1) Do the $K/\pi$ ratio $R_{K/\pi}$ and higher-dimensional second 
order factorial moment $F_2(M)$ analysis event by event simultaneousely.

(2) In order to get rid of the influence of transverse momentum 
conservation\r{5}, throw away the point(s) in ln-$F_2$ versus ln$M$ plot 
with $M_y<2$. Denote the first of the remaining points as $M_0$. Change the 
origin of coordinate system to this point:
  $$ x=\ln M-\ln M_0 \ , \ y=\ln F_2(M)-\ln F_2(M_0). \eqno(3)$$
Do a quadratic fit:
  $$ y = a x^2 + b x \eqno(4)$$
and use the fitting parameter $a$ as a characteristic of the degree of
upward-bending of ln-$F_2$ versus ln$M$.

(3) Draw the frequency of events as a function of their respective 
$R_{K/\pi}$ along with their upward-bending parameter $a$.
Plot the projection distributions of $R_{K/\pi}$ and $a$.

Then the 'QGP event sub-sample' can be selected by the following ways.

{\sl Method A:} If the distribution of $R_{K/\pi}$ has double peak or a peak
with a shoulder, then take the boundary of the two peaks or that of the peak
and the shoulder as $R_{K/\pi}^{\rm cut}$. Form the 'QGP event sub-sample' from 
the events with  $R_{K/\pi}>R_{K/\pi}^{\rm cut}$. 

{\sl Method B:} If the distribution of bending parameter $a$ has double peak or 
a peak with a shoulder, then take the boundary of the two peaks or that of the 
peak and the shoulder as $a^{\rm cut}$. Form the 'QGP event sub-sample' from 
the events with $a<a^{\rm cut}$. 

{\sl Method C:} In the scattering plot and contour plot of event frequency
as function of $R_{K/\pi}$ and $a$ try to find out the dense event
region at small $R_{K/\pi}$ and large $a$.  This dense region should mainly 
consist of non-QGP events. Identify the center of this region and draw an 
ellips with suitable long and short axes to cover this region as precisely 
as possible. Throwing away the events falling inside this ellips, a 'QGP 
event sub-sample' is obtained.
 
For illustration we have made a simple model\r{2} to do Monte Carlo simulation.
In this model a nucleus-nucleus collision event with $N$ elementary 
collisions is simulated by $N$ two-dimensional random cascading $\alpha$ 
models\footnote{The details of $\alpha$ model used here can be found 
in Ref.[7].}.  
Each $\alpha$ model corresponds to one of the $N$ elementary collisions. The
investigation region of the nuclear collision is taken to be $(0,1)\times(0,1)$.
The regions of the $N$ elementary collisions in direction $b$ ('transverse'
direction) are all (0,1), coinciding with that of the nuclear collision, while
the region in direction $a$ ('longitudinal' direction) for the $i$th elementary 
collision ($i=1,\dots,N$) is taken to be of width $1+\omega$ with the center 
placed randomly at $(\ (1-\omega)/2, (1+\omega)/2\ ),$ where $0<\omega \leq 1$ 
is a fixed parameter. The resulting particles from all the $N$ 
elementary $\alpha$ models falling in the region $(0,1)\times(0,1)$ are then 
superposed together to form the ``nucleus-nucleus collision event''. The model
parameters are taken to be $\alpha=0.5$ and $\omega=1$.

For a non-QGP  %% ordinary 
nucleus-nucleus collision event the number $N$ of elementary
collisions is taken to be 100, while for a QGP candidate event $N=1$. The 
number of charged particles in a single event is $n_{\rm ch}=1000$. 

In this way 2000 non-QGP events and the same number of QGP candidate events
are mixed to form a sample with $\rho_{\rm QGP}=50\%$. Taking randomly 1000
and 500 QGP candidate events from the 2000 events obtained above, mixing them 
with the 2000 non-QGP events, two samples with $\rho_{\rm QGP} = 33\%$ 
and $20\%$ respectively are obtained. 

The procedure 1--3 mentioned above are then done for the two samples with
$\rho_{\rm QGP}=50\%$ and 20\% and the Methods A, B and C are used to get 
the 'QGP sub-samples'. The results are shown in Fig.s 1--3 and Table I. 

%\newpage

{\bf Table I} $\rho_{\rm QGP}$ and percentage of lost QGP events in the
'QGP event sub-sample' obtained through three different methods.

\vskip0.5cm

%\noindent 
{\small \begin{tabular}{|c|c|c|c|c|c|}\hline
& \multicolumn{3}{|c|}{ $\rho_{\rm QGP}({\rm original}) = 50\% $}  
& \multicolumn{2}{|c|}{ $\rho_{\rm QGP}({\rm original}) = 20\% $} \\ \hline 
&  Method A & Method B & Method C & Method B & Method C \\ \hline
$\rho_{\rm QGP}$({\rm sub-sample}) & 95.5\% & 98.2\% & 98.6\% & 98.3\% & 94.7\%
\\ \hline
Lost QGP events & 33.6\% & 3.8\% & 2.5\% & 7.4\% & 2.8\%
\\ \hline
\end{tabular}}

\vskip0.5cm
%%%%%%%%%%%%%%%%%%%%%%%%%%%%%%%%%%%%%%%%%%%%%%%%%%%%%%%%%%%

The following observations can be made from the figures and Table I.

(1) As has been pointed out above, no evident shoulder can be seen  in the 
$R_{K/\pi}$ distribution when the percentage of 'QGP' events 
$\rho_{\rm QGP}\leq 30\%$ and the method (Method A) using $K/\pi$ ratio to 
select 'QGP' events failed in these cases. On the other hand, the method
using the upward-bending parameter $a$ of ln-FM versus ln$M$ (Method B)
and the method using $R_{K/\pi}$ and $a$ simultaneousely (Method C) can
be applied to very low $\rho_{\rm QGP}$ provided the number $N$ of elementary
nucleus-nucleus collision is high enough (greater than 100 for 
example)\footnote{We have done simulation with different number $N$ of 
elementary collisions. As $N$ increases, the $a$ distribution moves to larger 
$a$ and becomes more and more sharp. The Methods B and C can savely be 
applied when $N \geq 100$.}.  

(2) The percentage $\rho_{\rm QGP}$ of QGP events in the 'QGP candidate
sub-sample' is higher in Methods B and C than in Method A and the percentage
of lost QGP events in the sub-sample is lower in Methods B and C than in 
Method A. This shows that Methods B and C are more efficient in selecting
QGP events than Method A.

(3) The percentage $\rho_{\rm QGP}$ of QGP events in the 'QGP candidate
sub-sample' obtained by using 2-dimensional elliptical cut (Method C) is almost
equal to that using 1-dimensional $a$ cut (Method B), but less 'QGP' events 
are lost.

In this letter the method for selecting QGP candidate events from the 
whole relativistic heavy ion collision event sample has been discussed
in some detail. Basing on a simple geometrical consideration, the 
upward-bending of higher-dimensional factorial moment has been ultilized
to select QGP candidate events from the whole event sample.
The algorithm using the upward-bending parameter $a$ and the one using $a$
and $R_{K/\pi}$ simultaneousely have been shown to be more powerful and
more efficient than the one using $R_{K/\pi}$ alone. 

The algorithm proposed in this letter 
is primarily for the analysis of heavy ion collision data at CERN-SPS
energy. It can also be applied to the analysis of the coming
ultra-relativistic heavy ion collision data from Brookhaven-RHIC and
CERN-LHC experiments.
The comparison of the scattering plots of event frequency as function of
$R_{K/\pi}$ and $a$ for different energy and/or different centrality 
of heavy ion collisions might provide useful information about the 
thermalization and phase transition of the system.

\noindent {\bf Acknowledgement} The authors thank Wu Yuanfang anmd Liu Feng
for helpful discussions.

\newpage
\vskip1cm
\noindent{\Large Figure Captions}

\bigskip

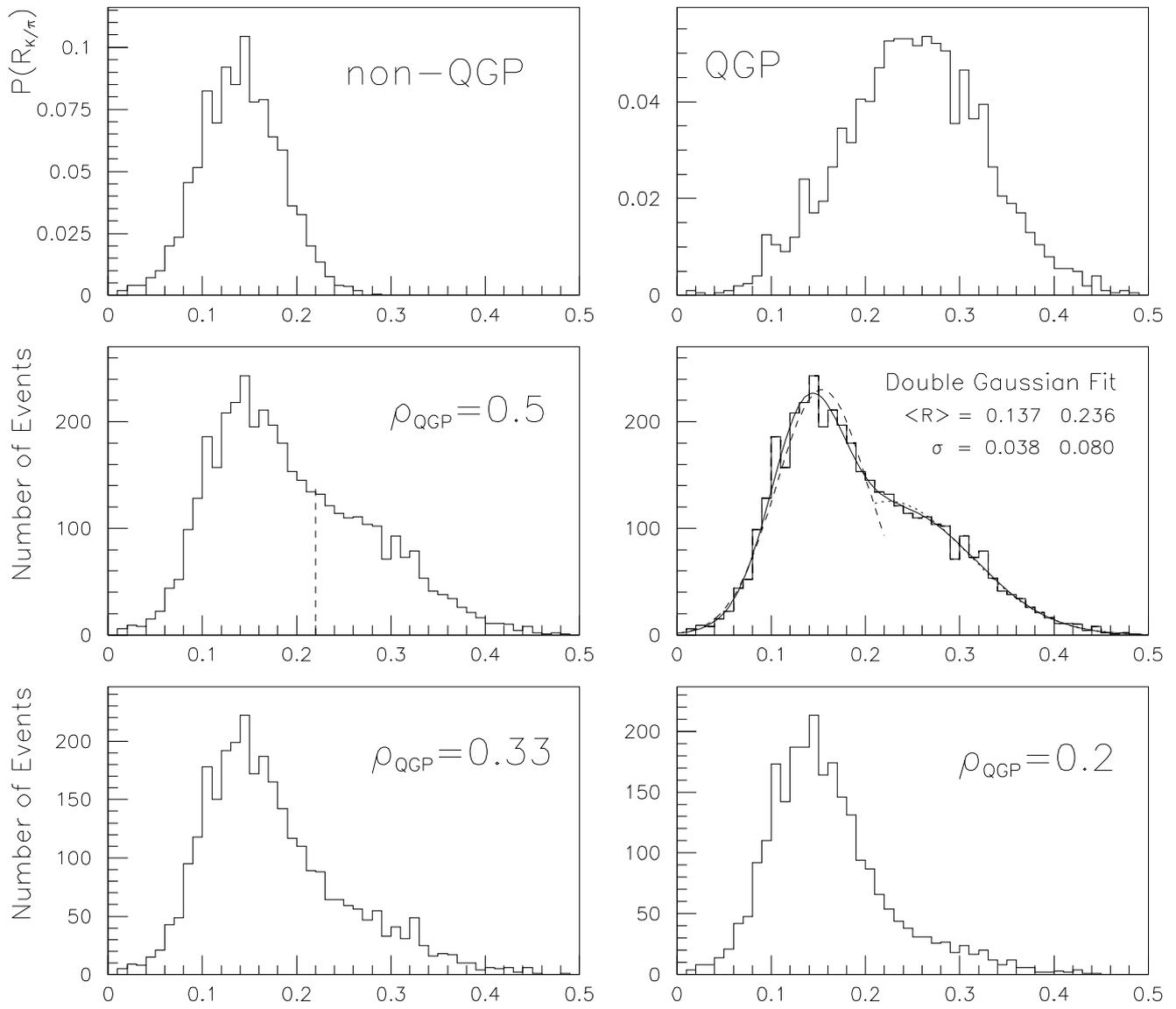
\noindent Fig. 1 \ Results of model calculation of '$K/\pi$ ratio' 
distribution for
'non-QGP', 'QGP' and mixed event sample. Dashed line is the $R(K/\pi)$ cut
for selecting 'QGP' events.

\bigskip

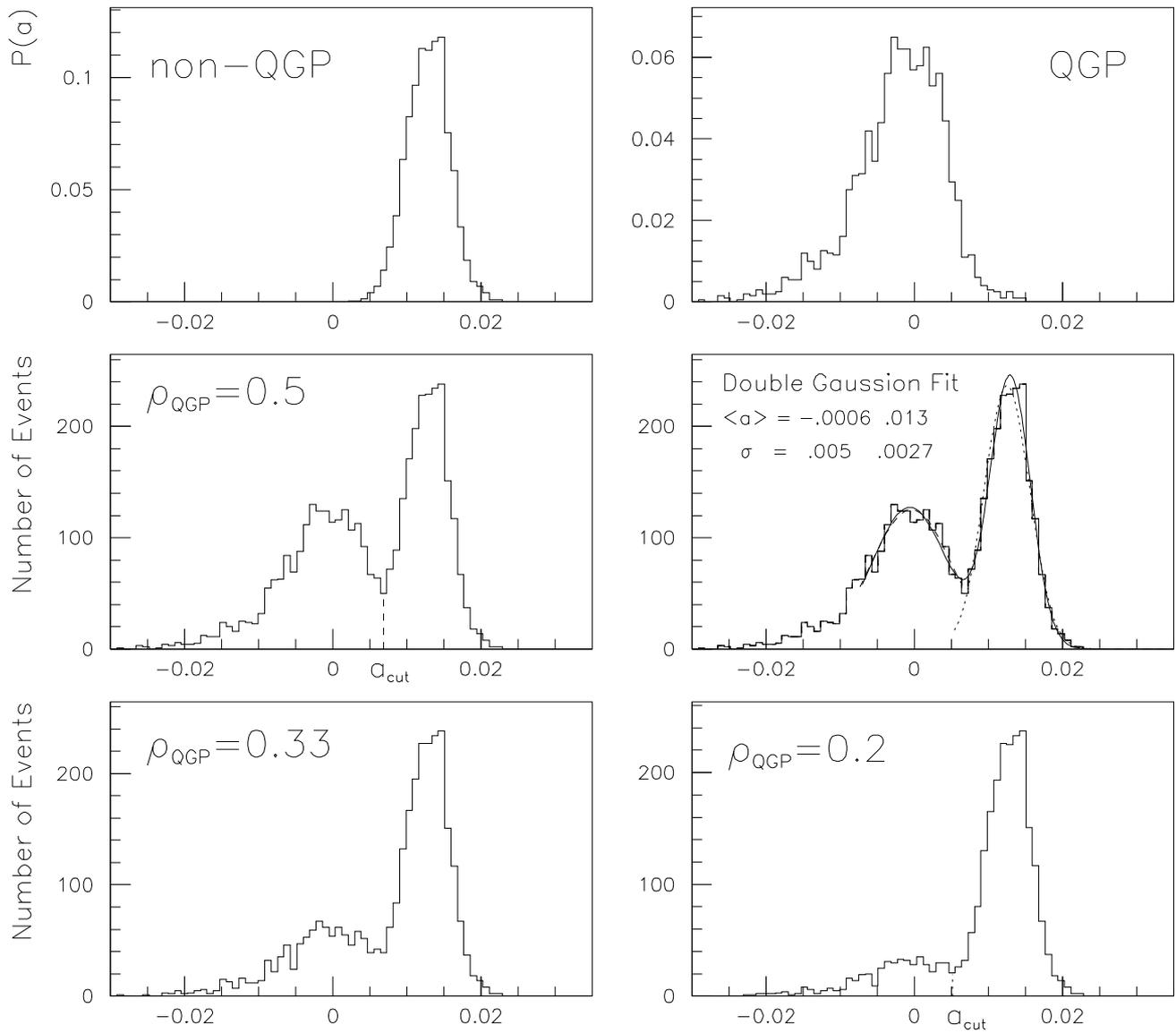
\noindent Fig. 2 \ Simulation results of the distribution of upward-bending
parameter $a$ of ln-FM versus ln$M$ for
'non-QGP', 'QGP' and mixed event sample. Dashed line is the $a$ cut
for selecting 'QGP' events.

\bigskip

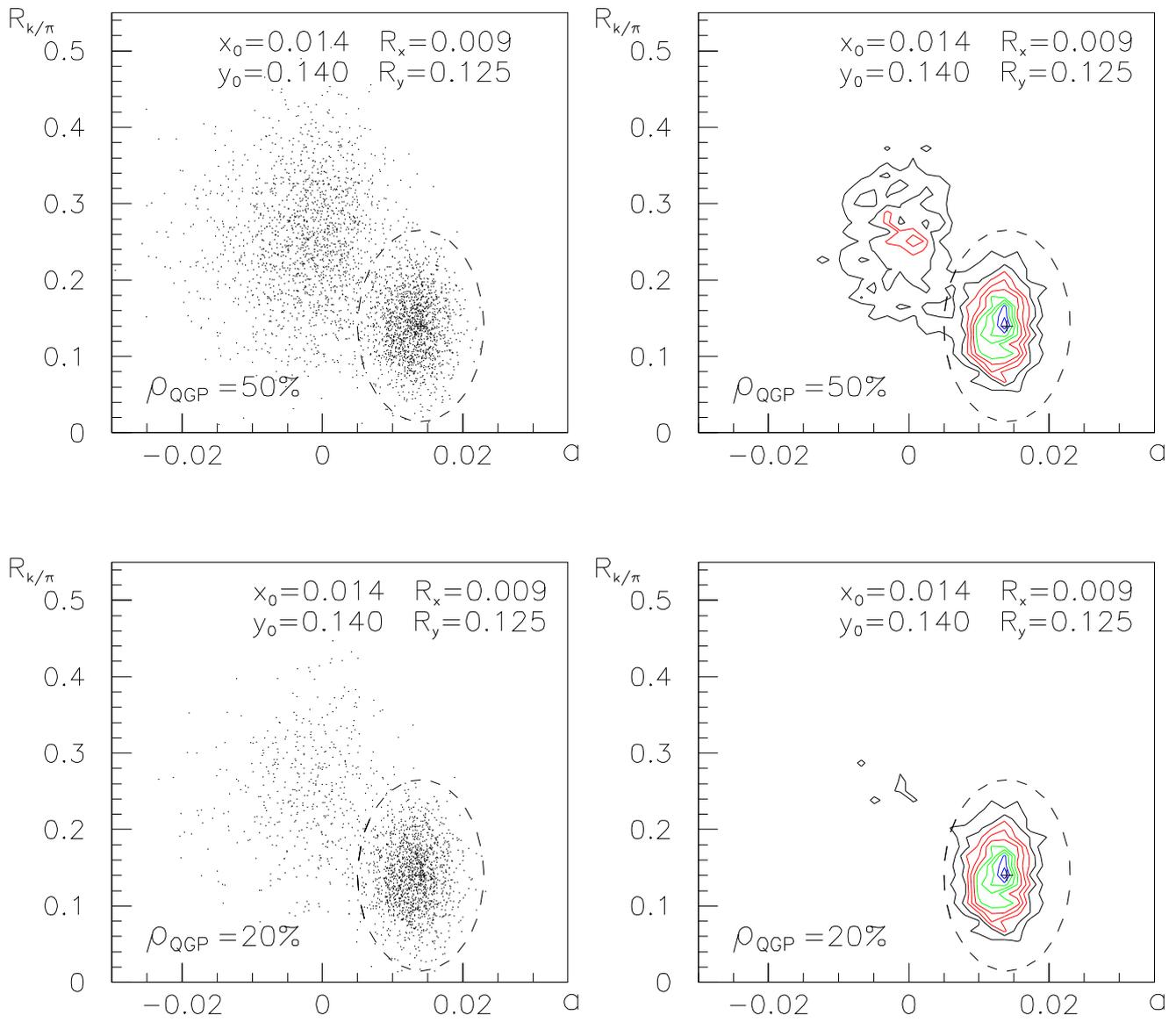
\noindent Fig. 3 \ The scattering plots and contour plots of event frequency
as function of $R_{K/\pi}$ and $a$ for mixed event sample from model 
simulation. The dashed ellips is the cut for selecting 'QGP' events.

\end{document}